\documentclass[aps,showpacs,preprintnumbers,amsmath, amssymb]{revtex4}

\usepackage{float}
\usepackage{graphics,epsfig}
\usepackage{graphicx}
\usepackage{dcolumn}
\usepackage{bm}

\begin{document}
\baselineskip=0.8 cm

\title{{\bf Dynamics of interacting dark energy model in Einstein and Loop Quantum Cosmology}}
\author{Songbai Chen}
\email{csb3752@163.com} \affiliation{ Institute of Physics and
Department of Physics, Hunan Normal University,  Changsha, Hunan
410081, P. R. China \\ Key Laboratory of Low Dimensional Quantum
Structures, \\ and Quantum Control of Ministry of Education, Hunan
Normal University, Changsha, Hunan 410081, P. R. China
\\ Department of Physics, Fudan University,
Shanghai 200433, P. R. China}

\author{Bin Wang}
\email{wangb@fudan.edu.cn} \affiliation{Department of Physics, Fudan
University, Shanghai 200433, P. R. China}

\author{Jiliang Jing}
\email{jljing@hunnu.edu.cn}
 \affiliation{ Institute of Physics and
Department of Physics, Hunan Normal University,  Changsha, Hunan
410081, P. R. China \\ Key Laboratory of Low Dimensional Quantum
Structures \\ and Quantum Control of Ministry of Education, Hunan
Normal University, Changsha, Hunan 410081, P. R. China}

\vspace*{0.2cm}
\begin{abstract}
\baselineskip=0.6 cm
\begin{center}
{\bf Abstract}
\end{center}

We investigate the background dynamics when dark energy is coupled
to dark matter in the universe described by Einstein cosmology and
Loop Quantum Cosmology. We introduce a new general form of dark
sector coupling, which presents us a more complicated dynamical
phase space. Differences in the phase space in obtaining the
accelerated scaling attractor in Einstein cosmology and Loop
Quantum Cosmology are disclosed.

\end{abstract}

 \maketitle
\newpage
\section{Introduction}

Our universe is undoubtedly undergoing an accelerated expansion
driven by a yet unknown dark energy (DE)\cite{A1,A2,A3,A4}. This
mysterious energy component occupies almost $70\%$ of the content of
the universe today. The leading interpretation of such a DE is a
cosmological constant with equation of state (EoS) $\omega = -1$.
Although this interpretation is consistent with observational data,
at the fundamental level it fails to be convincing. The vacuum
energy density is far below the value predicted by any sensible
quantum field theory, and it suffers the coincidence problem,
namely, ``why are the vacuum and matter energy densities of
precisely the same order today?".  To overcome the coincidence
problem, some sophisticated dynamical DE models relating the DE to
scalar fields have been put forward to replace the cosmological
constant \cite{T1}.

Considering that DE contributes a significant fraction of the
content of the universe, it is natural to look into its interaction
with the remaining fields of the Standard Model in the framework of
field theory. The possibility that DE and dark matter (DM) can
interact has got growing attention recently \cite{c1}-\cite{gf1}. It
has been argued that an appropriate interaction between DE and DM
can influence the perturbation dynamics and affect the lowest
multipoles of the CMB spectrum \cite{c4,c7}. Recently, it has been
shown that such a coupling can be inferred from the expansion
history of the Universe, as manifested in the supernova data
together with CMB and large-scale structure \cite{c8}. Signatures of
the interaction between DE and DM in the dynamics of galaxy clusters
has also been analyzed \cite{O1,O2}. It has been argued that the
coupling between DE and DM can provide a mechanism to alleviate the
coincidence problem and lead to an accelerated scaling attractor
solution with similar energy densities in the dark sectors today
\cite{c2,c6}.

A general interaction between DE and DM can be described in the
background by the balance equations
\begin{eqnarray}
&&\dot{\rho_x}+3H(1+\omega_x)\rho_x=-\Gamma,\nonumber\\
&&\dot{\rho_m}+3H\rho_m=\Gamma,\label{2}
\end{eqnarray}
where $\rho_m$ and $\rho_x$ correspond to the energy densities of
DM and DE, respectively. Here $\Gamma$ describes the coupling
between DE and DM. Since the nature of dark sectors remain
unknown, there is as yet no basis in fundamental theory for a
specific coupling in the dark sectors. All coupling models
discussed at the present moment are necessarily phenomenological
\cite{c1}. There are two criterions to determine whether some
models can be more physical justification than the others. One is
to confront observations. The other is to examine whether the
coupling can lead to accelerated scaling attractor
solutions\cite{pt}, which is a decisive way to achieve similar
energy densities in dark sectors and alleviate the coincidence
problem. In this work we introduce a new form of dark sector
coupling, $\Gamma=3cH\rho^{\alpha}_x\rho^{1-\alpha}_m$. This model
is more general than the coupling discussed in the literatures.
When $\alpha=0, 1$, it reduces to the cases with coupling between
dark sectors solely proportional to the energy densities of DM
\cite{dM1} and DE \cite{dE1}, respectively. These two limiting
cases have been examined thoroughly against observations and their
possibilities to alleviate coincidence problem have also been
discussed \cite{hb,gf}. Here we will investigate the background
dynamics when the DE is coupled to DM via this general interaction
form. We will show that the general coupling leads to a more
complicated dynamical phase space.

Besides the discussion of the dynamics of DE with our general
coupling to DM in the universe described by the Einstein theory, we
will also extend our investigation to the Loop Quantum Cosmology
(LQC). The LQC \cite{lc1,lc2,lc3} is the application of the Loop
Quantum Gravity \cite{lcg1,lcg2,lcg3} in the cosmological context,
which keeps the properties of non-perturbative and background
independent quantization of gravity. Recent investigations have
shown that the loop quantum effects can be very well described by an
effective modified Friedmann dynamics. There are two types of
modification to the Friedmann equation. The first one is based on
the modification to the behavior of inverse scale factor operator
below a critical scale factor $a_*$. Considering these modifications
one can obtain many interesting results including the replacement of
the classical big bang by a quantum bounce with desirable features
\cite{lcg31}, avoidance of many singularities \cite{lcg32}, easier
inflation \cite{lcg33}, and so on. However, the first type of
modification to Friedmann equation suffers from gauge dependence
which can not be cured and thus yields unphysical effects. The
second type of modification to Friedmann equation is discovered very
recently. It adds a $-\rho^2/\rho_c$ term in the standard Friedmann
equation which essentially encodes the discrete quantum geometric
nature of spacetime \cite{lc3,lg4,lg5}. When energy density of the
universe becomes of the same order of a critical density $\rho_c$,
this modification becomes dominant and the universe begins to bounce
and then oscillates forever. Thus the big bang singularity, the big
rip and other future singularities at semi-classical regime can be
avoided in LQC \cite{lc3,lg4,lg6,lg7,lg8}. Therefore by using the
second type of modification to Friedmann equation, the physically
appealing features of the first type are retained. For the universe
with a large scale factor, the first type of modification to the
effective Friedmann equation can be neglected and only the second
type of modification is important. Thus the dynamics of DE (phantom
\cite{lg7,phw1,phw2,phb1}, quintom and hessence \cite{h1}) have been
investigated recently in LQC on the basis of the second type of
modification. It finds that the dynamical properties of dark energy
models in LQC have different behaviors from those in the classical
Einstein cosmology. Here we will examine the background dynamics of
the LQC dominated by DE and DM where there is the general coupling
$\Gamma$ between dark sectors and compare the results with those in
Einstein cosmology.

The paper is organized as follows: in sections II and III, we
study the dynamics of the interacting dark energy model in
Einstein Cosmology and the LQC, respectively. In Sec.IV, we
present numerical pictures of dynamics in Einstein cosmology and
LQC. Our conclusions and discussions will be presented in the last
section.

\section{Dynamics of the interacting dark energy model in Einstein Cosmology}

In the Einstein theory, the universe is described by the standard
Friedmann equation
\begin{eqnarray}
H^2=\frac{\kappa}{3}\rho,\label{E1}
\end{eqnarray}
where $H$ is the Hubble parameter,  $\rho=\rho_m+\rho_x$ is the
total energy density and the constant $\kappa=8\pi G$. Since we are
concentrating on the late time accelerating universe, we have
neglected the radiation and baryons for simplicity.

Differentiating Eq.(\ref{E1}) and using the conservation law of
the total energy $\dot{\rho}+3H(\rho+p)=0$, we have
\begin{eqnarray}
\dot{H}=-\frac{\kappa}{2}(\rho+p).\label{H1}
\end{eqnarray}

To analyze the evolution of the dynamical system, we introduce the
dimensionless variables
\begin{eqnarray}
u\equiv\frac{\sqrt{\kappa\rho_x}}{\sqrt{3}H},\;\;\;\;\;\;v\equiv\frac{\sqrt{\kappa\rho_m}}{\sqrt{3}H},\;\;\;\;
\frac{d}{d\;N}=\frac{1}{H}\frac{d}{d\;t},\label{E4}
\end{eqnarray}
where $N\equiv\ln{a}$ is the number of $e$-folding to represent
the cosmological time. Using the above definitions, the Hubble
equations can be rewritten as
\begin{eqnarray}
u^2+v^2=1,
\end{eqnarray}
and
\begin{eqnarray}
\frac{\dot{H}}{H^2}=-\frac{3}{2}\bigg[1+\frac{\omega_xu^2}{u^2+v^2}\bigg]=-\frac{3}{2}(1+\omega_xu^2).\label{EHH}
\end{eqnarray}
The effective total EOS $\omega_{tot}$ is given by
\begin{eqnarray}
\omega_{tot}=\frac{\omega_x\rho_x}{\rho_x+\rho_m}=\frac{\omega_xu^2}{u^2+v^2}=\omega_xu^2.\label{st1}
\end{eqnarray}
Using dimensionless variables, the dynamical equations of the
system can be expressed as
\begin{eqnarray}
u'&=&\frac{3u}{2}\bigg[\omega_x(u^2-1)-c\bigg(\frac{u^2}{v^2}\bigg)^{\alpha-1}\bigg],\nonumber\\
v'&=&\frac{3v}{2}\bigg[\omega_xu^2+c\bigg(\frac{u^2}{v^2}\bigg)^{\alpha}\bigg],
\end{eqnarray}
where the prime denotes a derivative with respect to $N$. The
critical points $u_c, v_c$ satisfy $u'=0$ and $v'=0$. In order to
study the stability of the critical points, we expand about the
critical points $u=u_c+\delta u, v=v_c+\delta v$ and linearize the
above equations near the critical points so that
\begin{eqnarray}
\delta
u'&=&\frac{3}{2}\bigg[3\omega_xu_c^2-1-(2\alpha-1)c\bigg(\frac{u_c}{v_c}\bigg)^{2\alpha-2}\bigg]\delta
u+\bigg[3c(\alpha-1)\bigg(\frac{u_c}{v_c}\bigg)^{2\alpha-1}\bigg]\delta
v,\nonumber\\
\delta
v'&=&\bigg[3\omega_xu_cv_c+3c\alpha\bigg(\frac{u_c}{v_c}\bigg)^{2\alpha-1}\bigg]\delta
u+\frac{3}{2}\bigg[\omega_xu^2_c-(2\alpha-1)c\bigg(\frac{u_c}{v_c}\bigg)^{2\alpha}\bigg]\delta
v.
\end{eqnarray}
The eigenvalues of the matrix of coefficients of the above
equations encode the behavior of the dynamical system near the
critical points.

In general, for an arbitrary $\alpha$ it is difficult to obtain
the analytical forms of the critical points. Here we only consider
some specific values of $\alpha$ and examine the dynamics of DE
with an interaction with DM.

\subsection{Case I: \;\ $\alpha=0$}

When $\alpha=0$, the interaction form reduces to the simple coupling
between DE and DM in proportional to the energy density of DM. This
simple interaction form has been confronted to observations and its
possibility to alleviate the coincidence problem has also been
examined \cite{hb}. From the dynamical system equations, we can
obtain two critical points for the specific model, namely:
\begin{eqnarray}
&&\bullet \;\;\text{Point}\; A_0: \;( 1,\;\;0),\nonumber \\ &&
\bullet \;\;\text{Point}\; B_0: \; \bigg(\;
\sqrt{-\frac{c}{\omega_x}},\;\;\sqrt{1+\frac{c}{\omega_x}}\;\bigg).
\end{eqnarray}
The eigenvalues of the coefficient matrix of the linearized
equations around these critical points can be expressed
respectively as
\begin{eqnarray}
&&\bullet \;\;\text{Point} \;A_0: \;\;\;\;\;
\lambda_1=3\omega_x,\;\;\;\;\;
\lambda_2=\frac{3}{2}(c+\omega_x),\nonumber\\
&&\bullet \;\;\text{Point}\; B_0: \;\;\;\;\;
\lambda_1=-3c,\;\;\;\;\; \lambda_2=-3(c+\omega_x).
\end{eqnarray}
For point $A_0$, when $c<-\omega_x$, both eigenvalues $\lambda_1$
and $\lambda_2$ are negative, which indicates that $A_0$ is a stable
point. From Eq.(\ref{st1}), we learn that the effective total EOS at
point $A_0$ is $\omega_{tot}=\omega_x$. Therefore we obtain
$\ddot{a}\propto-(1+3\omega_{x})\;t^{\frac{2}{3(1+\omega_{x})}-2}$
and $\rho\propto a^{-3(1+\omega_{x})}$. It means that point $A_0$ is
an accelerated scaling solution as $\omega_x<-1/3$ and there is
singularity in the finite future as $\omega_x<-1$. When
$c>-\omega_x$, the sign of $\lambda_1$ is always opposite to the
sign of $\lambda_2$, which leads $A_0$ to a saddle point. For point
$B_0$, the critical point $v_c$ exists only provided that
$c\leq-\omega_x$, which leads the sign of $\lambda_1$ always
opposite to that of $\lambda_2$. Thus point $B_0$ is a saddle point.

\subsection{Case II: \;\ $\alpha=1$}

In this limiting case the coupling between dark sectors is in
proportional to the energy density of DE. This simple coupling has
been examined using observational data and its effect to alleviate
the coincidence problem has been discussed \cite{hb}.  One can
obtain two critical points in this dynamical system:
\begin{eqnarray}
&&\bullet \;\;\text{Point}\; A_1: \;( 0,\;\;1),\nonumber \\ &&
\bullet \;\;\text{Point}\; B_1: \; \bigg(\;
\sqrt{1+\frac{c}{\omega_x}},\;\;\sqrt{-\frac{c}{\omega_x}}\;\bigg).
\end{eqnarray}
The eigenvalues of the coefficient matrix of the linearized
equations are
\begin{eqnarray}
&&\bullet \;\;\text{Point} \;A_1: \;\;\;\;\; \lambda_1=0,\;\;\;\;\;
\lambda_2=-\frac{3}{2}(c+\omega_x),\nonumber\\
&&\bullet \;\;\text{Point}\; B_1: \;\;\;\;\;
\lambda_1=3(c+\omega_x),\;\;\;\;\; \lambda_2=3(c+\omega_x).
\end{eqnarray}
It is easy to examine that the critical point $A_1$ is not a stable
point, while $B_1$ is stable when $c\leq-\omega_x$. The total
effective EOS at point $B_1$ reads $\omega_{tot}=c+\omega_x$. When
$\omega_x\leq-1/3-c$, point $B_1$ can be an accelerated scaling
solution. From Eq.(\ref{EHH}) we find
$H=\frac{2}{3(1+\omega_x+c)(t_0-t)},\;
\dot{H}=-\frac{2}{3(1+\omega_x+c)(t_0-t)^2}$. Thus as
$\omega_x\leq-1-c$ the universe will undergo super-accelerated
expansion ($\dot{H}>0$) and end in the big rip.

\subsection{Case III: \;\ $\alpha=\frac{1}{2}$}

Solving equations $u'=0$ and $v'=0$, one can obtain two critical
points of the dynamical system:
\begin{eqnarray}
&&\bullet \;\;\text{Point}\; A_2: \;\bigg(
\sqrt{\frac{1}{2}+\frac{1}{2}\sqrt{1-\frac{4c^2}{\omega^2_x}}},
\;\;\sqrt{\frac{1}{2}-\frac{1}{2}\sqrt{1-\frac{4c^2}{\omega^2_x}}}\bigg),\nonumber \\
&& \bullet \;\;\text{Point}\; B_2: \; \bigg(\;
\sqrt{\frac{1}{2}-\frac{1}{2}\sqrt{1-\frac{4c^2}{\omega^2_x}}},
\;\;\sqrt{\frac{1}{2}+\frac{1}{2}\sqrt{1-\frac{4c^2}{\omega^2_x}}}\;\bigg).
\end{eqnarray}
Through analysis of the eigenvalues of the coefficient matrix, we
find that point $A_2$ is stable when $c<-\frac{\omega_x}{2}$, while
point $B_2$ is a saddle point. At point $A_2$, the total effective
EOS is $\omega_{tot}=(\omega_x-\sqrt{\omega^2_x-4c^2})/2$, which
shows that if $c>\frac{1}{3}$ and $\omega_x<-2c$ or $c<\frac{1}{3}$
and $-\frac{2}{3}<\omega_x<-\frac{1}{3}-3c^2$, we have
$\omega_{tot}<-\frac{1}{3}$ so that $A_2$ is an accelerated scaling
attractor. When $\omega_x<\frac{2}{3}(1-2\sqrt{1+3c^2})$, one
obtains $\omega_{tot}<-1$, $\dot{H}>0$ and finds there is a future
singularity in this case.

\subsection{Case IV: \;\ $\alpha=-1$}

In this case the system has two critical points:
\begin{eqnarray}
&&\bullet \;\;\text{Point}\; A_3: \;( 1,
\;\;\;0),\nonumber \\
&& \bullet \;\;\text{Point}\; B_3: \; \bigg(\;
\sqrt{\frac{c-\sqrt{c^2-4\omega_xc}}{2\omega_x}},
\;\;\sqrt{1-\frac{c-\sqrt{c^2-4\omega_xc}}{2\omega_x}}\;\bigg).
\end{eqnarray}
In term of the signs of the eigenvalues of the coefficient matrix,
we find that point $A_3$ is stable when $\omega_x<0$, however, point
$B_3$ is a saddle point. From Eq.(\ref{st1}), we learn that point
$A_3$ can be an accelerated scaling attractor provided that
$\omega_{x}<-\frac{1}{3}$ and there is a future singularity as
$\omega_{x}<-1$.

\subsection{Case IV: \;\ $\alpha=2$}

When $\alpha=2$, the system has two critical points:
\begin{eqnarray}
&&\bullet \;\;\text{Point}\; A_4: \;(0,
\;\;\;1),\nonumber \\
&& \bullet \;\;\text{Point}\; B_4: \;
\bigg(\;\sqrt{1-\frac{c-\sqrt{c^2-4\omega_xc}}{2\omega_x}},
\;\;\;\sqrt{\frac{c-\sqrt{c^2-4\omega_xc}}{2\omega_x}}\bigg).
\end{eqnarray}
Similarly, we find that point $A_4$ is unstable, while point $B_4$
is stable for all $\omega_x<0$.  The total effective EOS at point
$B_4$ is $\omega_{tot}=\omega_x-(c-\sqrt{c^2-4\omega_xc})/2$. When
$\omega_x<-\frac{1}{3}(1+\sqrt{3c})$, we have
$\omega_{tot}<-\frac{1}{3}$ and $B_4$ can be an accelerated scaling
solution. Moreover, we find that $\omega_{tot}>-1$ as
$\omega_x>-1-\sqrt{c}$ and there is no future singularity in this
case. But as $\omega_x<-1-\sqrt{c}$ one can obtain $\omega_{tot}<-1$
and a future singularity is inevitable. Thus in the Einstein
cosmology the presence of the coupling terms can not remove the
singularity entirely.

\section{Dynamics of the interacting dark energy model in LQC}

In this section we are going to extend our discussion to the LQC.
The loop quantum effect modifies the Friedmann equation into
\cite{lc3,lg4,lg5,lg51}
\begin{eqnarray}
H^2=\frac{\kappa}{3}\rho\bigg(1-\frac{\rho}{\rho_c}\bigg),\label{1}
\end{eqnarray}
where $\rho_c\equiv \frac{\sqrt{3}}{16\pi^2\gamma^3G^2\hbar}$ is the
critical loop quantum density and $\gamma$ is the Barbero-Immirzi
parameter. Let us note here it has been suggested that
$\gamma\approx 0.2375$ by the black hole thermodynamics in LQG
\cite{lg6}.

Differentiating Eq.(\ref{1}) and using the conservation equation of
the total energy density $\dot{\rho}+3H(\rho+p)=0$, where
$\rho=\rho_x+\rho_m$, one can obtain
\begin{eqnarray}
\dot{H}=-\frac{\kappa}{2}(\rho+p)\bigg(1-\frac{2\rho}{\rho_c}\bigg).
\end{eqnarray}
Adopting the dimensionless variables defined in (\ref{E4}), the
evolution of the Hubble parameter in LQC becomes
\begin{eqnarray}
(u^2+v^2)\bigg(1-3H^2\frac{u^2+v^2}{\rho_c}\bigg)=1,\label{1s}
\end{eqnarray}
and
\begin{eqnarray}
\frac{\dot{H}}{H^2}=-\frac{3}{2}(2-u^2-v^2)\bigg[1+\frac{\omega_xu^2}{u^2+v^2}\bigg].\label{EHH2}
\end{eqnarray}
The total effective EOS $\omega_{tot}$ in LQC is given by
\begin{eqnarray}
\omega_{tot}=\frac{\omega_x\rho_x}{\rho_x+\rho_m}=\frac{\omega_xu^2}{u^2+v^2}.\label{st2}
\end{eqnarray}
The dynamical system can be expressed as
\begin{eqnarray}
u'&=&\frac{3u}{2}\bigg[-1-\omega_x-c\bigg(\frac{u}{v}\bigg)^{2\alpha-2}
+\frac{[(1+\omega_x)u^2+v^2](2-u^2-v^2)}{u^2+v^2}\bigg],\nonumber\\
v'&=&\frac{3v}{2}\bigg[-1+c\bigg(\frac{u}{v}\bigg)^{2\alpha}
+\frac{[(1+\omega_x)u^2+v^2](2-u^2-v^2)}{u^2+v^2}\bigg],\label{L4}
\end{eqnarray}

Obtaining the critical points and linearizing the system near
them, we can study the stability of critical points by analyzing
the first-order differential equations
\begin{eqnarray}
\delta
u'&=&\frac{3}{2}\bigg[-3(\omega_x+1)u^2-v^2+1-\omega_x+\frac{2u^2(u^2+3v^2)}{(u^2+v^2)^2}
-(2\alpha-1)c\bigg(\frac{u}{v}\bigg)^{2\alpha-2}\bigg]\delta
u\nonumber\\
&-&3\bigg[uv+\frac{2\omega_xu^3v}{(u^2+v^2)^2}-(\alpha-1)c\bigg(\frac{u}{v}\bigg)^{2\alpha-1}\bigg]\delta
v,\nonumber\\
\delta
v'&=&3\bigg[-(\omega_x+1)uv+\frac{2\omega_xuv^3}{(u^2+v^2)^2}+\alpha
c\bigg(\frac{u}{v}\bigg)^{2\alpha-1}\bigg]\delta
u\nonumber\\
&+&\frac{3}{2}\bigg[1-\omega_x-u^2-3v^2+\frac{2\omega_xu^2(u^2-v^2)}{(u^2+v^2)^2}-
(2\alpha-1)c\bigg(\frac{u}{v}\bigg)^{2\alpha}\bigg]\delta v.
\end{eqnarray}
Solving the eigenvalues of the coefficient matrix of the above
equations, we can know the behavior of the dynamical system near
the critical points. Comparing with the Einstein theory, we find
that the same critical points will have different eigenvalues in
the coefficient matrix in LQC. This means that the dynamical
property of the system in the LQC is different from that in the
Einstein cosmology.

As did in section II, here we will also focus on some specific
cases for simplicity, such as $\alpha=0,\; 1,\; \frac{1}{2}, \;-1,
\;2$. For these specific cases, the critical points in the LQC are
the same as those in the Einstein cosmology.

\subsection{Case I: \;\ $\alpha=0$}

In this limiting case, the critical points of the dynamical system
are
\begin{eqnarray}
&&\bullet \;\;\text{Point}\; A_0: \;( 1,\;\;0),\nonumber \\ &&
\bullet \;\;\text{Point}\; B_0: \; \bigg(\;
\sqrt{-\frac{c}{\omega_x}},\;\;\sqrt{1+\frac{c}{\omega_x}}\;\bigg).
\end{eqnarray}
The eigenvalues of the linearized equation around the critical
points read
\begin{eqnarray}
&&\bullet \;\;\text{Point} \;A_0: \;\;\;\;\;
\lambda_1=-3(1+\omega_x),\;\;\;\;\;
\lambda_2=\frac{3}{2}(c+\omega_x),\nonumber\\
&&\bullet \;\;\text{Point}\; B_0: \;\;\;\;\;
\lambda_1=3(c-1),\;\;\;\;\; \lambda_2=-3(c+\omega_x).
\end{eqnarray}
It is easy to see that for the DE with $\omega_x>-1$ point $A_0$ is
stable if $c<-\omega_x$. The effective total EOS
$\omega_{tot}=\omega_x$, thus the critical point $A_0$ is an
accelerated scaling solution without a future singularity for
$-1<\omega_x<-1/3$. When the DE is phantom like $\omega_x<-1$, $A_0$
is a saddle point. This is different from that in the Einstein
theory, where $A_0$ is still stable for the phantom DE provided that
$c<-\omega_x$. Point $B_0$ is a saddle point if $c<1$ and is
unstable when $c>1$. In the Einstein theory, $B_0$ is always a
saddle points. This means that the presence of the term
$-\frac{\rho}{\rho_c}$ in the Friedmann equation due to the quantum
correction changes the dynamical properties of autonomous system.
Moreover, when the value of $\omega_x$ goes beyond the ranges of
$-1<\omega_x<-1/3$, the quantum bounce originated from the term
$-\frac{\rho}{\rho_c}$ will leads to the avoidance of the future
singularity.

\subsection{Case II: \;\ $\alpha=1$}

The critical points are the same as those in the Einstein theory
\begin{eqnarray}
&&\bullet \;\;\text{Point}\; A_1: \;( 0,\;\;1),\nonumber \\ &&
\bullet \;\;\text{Point}\; B_1: \; \bigg(\;
\sqrt{1+\frac{c}{\omega_x}},\;\;\sqrt{-\frac{c}{\omega_x}}\;\bigg).
\end{eqnarray}
However, the eigenvalues of the coefficient matrix of the
linearized equations near the critical points become
\begin{eqnarray}
&&\bullet \;\;\text{Point} \;A_1: \;\;\;\;\; \lambda_1=-3,\;\;\;\;\;
\lambda_2=-\frac{3}{2}(c+\omega_x),\nonumber\\
&&\bullet \;\;\text{Point}\; B_1: \;\;\;\;\;
\lambda_1=3(c+\omega_x),\;\;\;\;\; \lambda_2=-3(1+c+\omega_x).
\end{eqnarray}
Point $A_1$ is a stable point when $c>-\omega_x$. From Eq.
(\ref{EHH2}) we have $\omega_{tot}=0$ at the point $A_1$, which
means that in this model our universe described by LQC will enter DM
dominated era and there is no singularity in the finite future.
However, in the Einstein theory $A_1$ is unstable. Point $B_1$ can
be stable provided that $-(1+\omega_x)<c<-\omega_x$. This region of
$c$ to keep $B_1$ stable is smaller than that in the Einstein
theory. At point $B_1$, we have $\omega_{tot}=c+\omega_x$, which
shows that when $-1-c<\omega_x<-\frac{1}{3}-c$, we have
$-1<\omega_{tot}<-\frac{1}{3}$, so that point $B_1$ corresponds to
an accelerated attractor without a future singularity.

\subsection{Case III: \;\ $\alpha=\frac{1}{2}$}

The critical points can be found at
\begin{eqnarray}
&&\bullet \;\;\text{Point}\; A_2: \;\bigg(
\sqrt{\frac{1}{2}+\frac{1}{2}\sqrt{1-\frac{4c^2}{\omega^2_x}}},
\;\;\sqrt{\frac{1}{2}-\frac{1}{2}\sqrt{1-\frac{4c^2}{\omega^2_x}}}\bigg),\nonumber \\
&& \bullet \;\;\text{Point}\; B_2: \; \bigg(\;
\sqrt{\frac{1}{2}-\frac{1}{2}\sqrt{1-\frac{4c^2}{\omega^2_x}}},
\;\;\sqrt{\frac{1}{2}+\frac{1}{2}\sqrt{1-\frac{4c^2}{\omega^2_x}}}\;\bigg).
\end{eqnarray}
Analyzing the stability, we find that when DE is of quintessence
type, point $A_2$ is stable provided that $c<-\frac{\omega_x}{2}$.
When the DE is of phantom type, $A_2$ can be stable only when
$\sqrt{-(1+\omega_x)}<c<-\frac{\omega_x}{2}$ and $\omega_x>-2$. From
the total effective EOS, we learn that when $c>\frac{1}{3}$ and
$-1-c^2<\omega_x<-2c$ or $c<\frac{1}{3}$ and
$-1-c^2<\omega_x<-\frac{1}{3}-3c^2$, point $A_2$ is an accelerated
attractor. Similarly, when the value of $\omega_x$ goes beyond the
ranges above, the effects from the term $-\frac{\rho}{\rho_c}$ will
make the future singularity disappear. Point $B_2$ is a saddle
point, which agrees with that found in the Einstein theory.

\subsection{Case IV: \;\ $\alpha=-1$}

Critical points of the system read
\begin{eqnarray}
&&\bullet \;\;\text{Point}\; A_3: \;( 1,
\;\;\;0),\nonumber \\
&& \bullet \;\;\text{Point}\; B_3: \; \bigg(\;
\sqrt{\frac{c-\sqrt{c^2-4\omega_xc}}{2\omega_x}},
\;\;\sqrt{1-\frac{c-\sqrt{c^2-4\omega_xc}}{2\omega_x}}\;\bigg).
\end{eqnarray}
For the quintessence type DE, point $A_3$ is a stable point. Since
$\omega_{tot}=\omega_x$, $A_3$ is an accelerated scaling
attractor. For the phantom type DE, $A_3$ is a saddle point when
$c<-\frac{1}{1+\omega_x}$ and is unstable when
$c>-\frac{1}{1+\omega_x}$. Point $B_3$ is a saddle point always.

\subsection{Case IV: \;\ $\alpha=2$}

When $\alpha=2$, the system has two critical points:
\begin{eqnarray}
&&\bullet \;\;\text{Point}\; A_4: \;(0,
\;\;\;1),\nonumber \\
&& \bullet \;\;\text{Point}\; B_4: \;
\bigg(\;\sqrt{1-\frac{c-\sqrt{c^2-4\omega_xc}}{2\omega_x}},
\;\;\;\sqrt{\frac{c-\sqrt{c^2-4\omega_xc}}{2\omega_x}}\bigg).
\end{eqnarray}
Similarly, we find that point $A_4$ is a saddle point as that in the
Einstein theory. When the DE is of quintessence type, point $B_4$ is
stable and when $-1<\omega_x<-\frac{1}{3}(1+\sqrt{3c})$ $B_4$ is an
accelerated scaling attractor. When the DE is of phantom type, $B_4$
can be stable only when $c>(1+\omega_x)^2$. When
$-(1+\sqrt{c})<\omega_x<-\frac{1}{3}(1+\sqrt{3c})$, $B_4$ can be an
accelerated scaling solution as well. Although $\omega_{tot}<-1$ as
$\omega_x<-1-\sqrt{c}$, the loop quantum effects will cancel off the
future singularity in the evolution of the universe. Thus in the LQC
the big rip can be removed entirely.

\section{numerical results}
In this section we confirm numerically the complicated stability
conditions for critical points obtained above. For the new general
form of the interaction between DE and DM, $\Gamma=3cH
\rho^{\alpha}_x\rho^{1-\alpha}_m$, we have more complicated
dynamical phase spaces. We find that the position of the critical
point and its stability depend not only on the coupling constant
$c$, the DE EOS $\omega_x$ and the exponent $\alpha$ in the
coupling, but also on the theory to describe the universe. In figure
(1), we show the stable regions in the parameter space $(c,
\omega_x)$ by choosing $\alpha=0.5$ and $2$. In Einstein cosmology
the critical points $A_2$ ($\alpha=0.5$) and $B_4$ ($\alpha=2$) are
late time attractor in the region $I+II$. However in LQC, $A_2$ and
$B_4$ are late time attractors only in the region $II$. In LQC we
see that the region of the location of the accelerated scaling
attractor has been reduced compared to the Einstein theory. This
holds true for other values of $\alpha$.

In figure (2), we plot the numerical result to illustrate the
phase space trajectories for our coupling model with chosen $c$
and $\omega_x$ in the stable region and selected $\alpha$ in the
Einstein cosmology and the LQC. Since the LQC has  the same
critical points as that in the Einstein cosmology, we see that for
fixed $\alpha$, all trajectories with different initial conditions
in the Einstein cosmology and LQC converge to the same final state
determined by parameters $c$ and $\omega_x$. This means that our
universe will enter an era with similar energy densities of DE and
DM.

In figure (3), we show the evolution of the effective EOS
$\omega_{tot}=\omega_x\rho_x/(\rho_m+\rho_x)$ in the stable
region. We see that in the final state the total state parameter
$\omega_{tot}$ tends to a constant, which is determined by values
of $\rho_{x_c}$, $\rho_{m_c}$, $\omega_x$ and $\alpha$. For
selected values of $\omega_x, c$ to be within a certain range
discussed in sections II and III for different values of $\alpha$,
such as $\omega_x=-0.6$ and $c=0.18$, we can have stable critical
point and meanwhile we can get the total effective EOS
$\omega_{tot}<-\frac{1}{3}$ in the final state as displayed in
figure (3). Our universe will enter a final state with a constant
energy ratio between DE and DM and accelerate forever for all
chosen $\alpha$s. However when values of $\omega_x, c$ are beyond
the range discussed in Einstein cosmology and LQC, our universe
will enter a decelerated expansion.

In figure (4) we exhibit the evolution of $\rho(t)$ for chosen
$\alpha=-1$, $\rho_c=0.82$ and parameters $(\omega_x, c)$ in the
unstable region in LQC. We find that the universe finally enters an
oscillating regime in the LQC. The oscillating frequencies of
$\rho(t)$ depend on the coupling constant $c$ and DE EOS $\omega_x$.
This oscillation behavior makes the universe experience bouncing,
which can avoid singularity faced in the usual Einstein cosmology.
This property also holds for other values of $\alpha$ in LQC.

\section{Conclusions and discussions}
In this paper we have studied the background dynamics when DE is
modelled coupling with DM via a new general form $3cH
\rho^{\alpha}_x\rho^{1-\alpha}_m$ in Einstein cosmology and LQC.
For selected values of $\alpha$, we have examined stability
behaviors of critical points and found accelerated scaling
solutions to account for the similar energy densities in dark
sectors today. In LQC, the parameter space for the existence of
the accelerated scaling attractor is found smaller than that in
the Einstein cosmology. In the unstable region, the universe
described by the LQC will enter an oscillatory regime which can
help to avoid the singularity usually met in Einstein cosmology.

The background dynamics for the new general form of dark sector
coupling leads to more complicated features in dynamical phase
space. In order to confront this model with observations, the
cosmological perturbations with this coupling form need to be
disclosed. Some discussions in this direction has been addressed
recently in \cite{pp}.

\begin{acknowledgments}

We are thankful to Dr. P. X. Wu for his useful discussion. This work
was partially supported by NNSF of China, Shanghai Education
Commission and Shanghai Science and Technology Commission. S. B.
Chen's work was partially supported by the National Natural Science
Foundation of China under Grant No.10875041; the Scientific Research
Fund of Hunan Provincial Education Department Grant No.07B043 and
the construct program of key disciplines in Hunan Province. J. L.
Jing's work was partially supported by the National Natural Science
Foundation of China under Grant No.10675045; the FANEDD under Grant
No. 200317; and the Hunan Provincial Natural Science Foundation of
China under Grant No.08JJ3010.
\end{acknowledgments}
\begin{figure}[ht]
\begin{center}
\includegraphics[width=6cm]{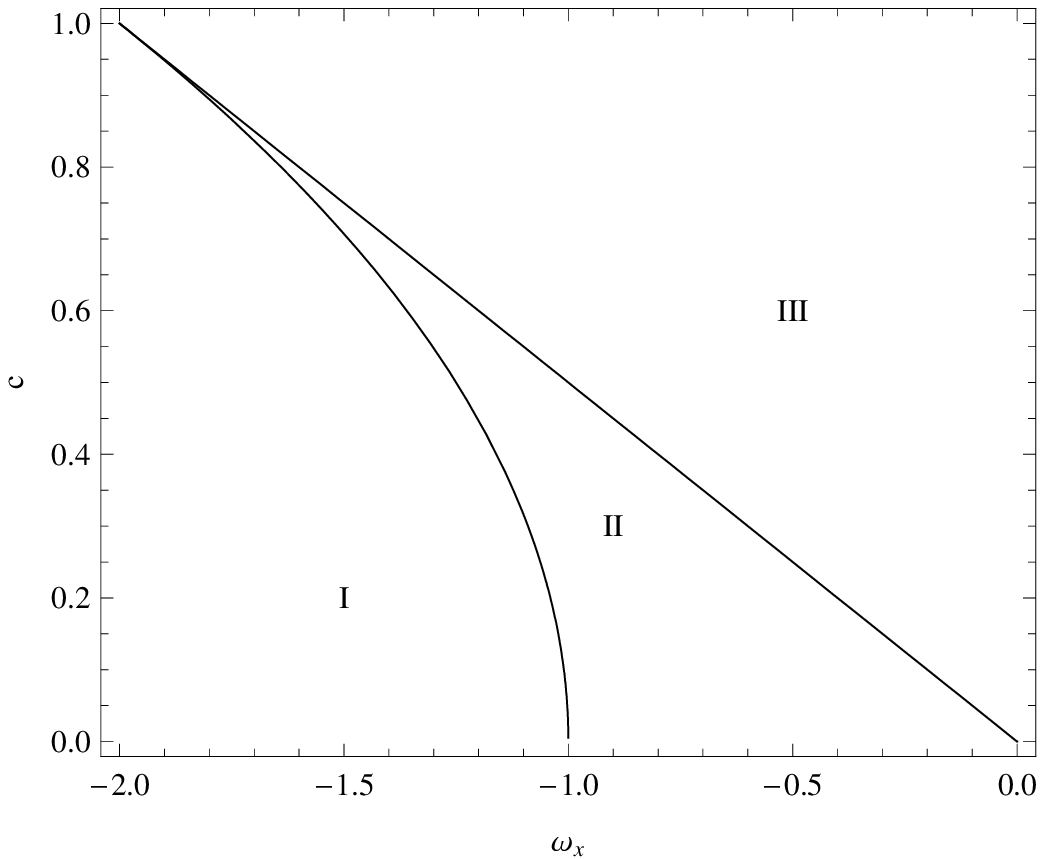}\;\;\;\;\;\includegraphics[width=6cm]{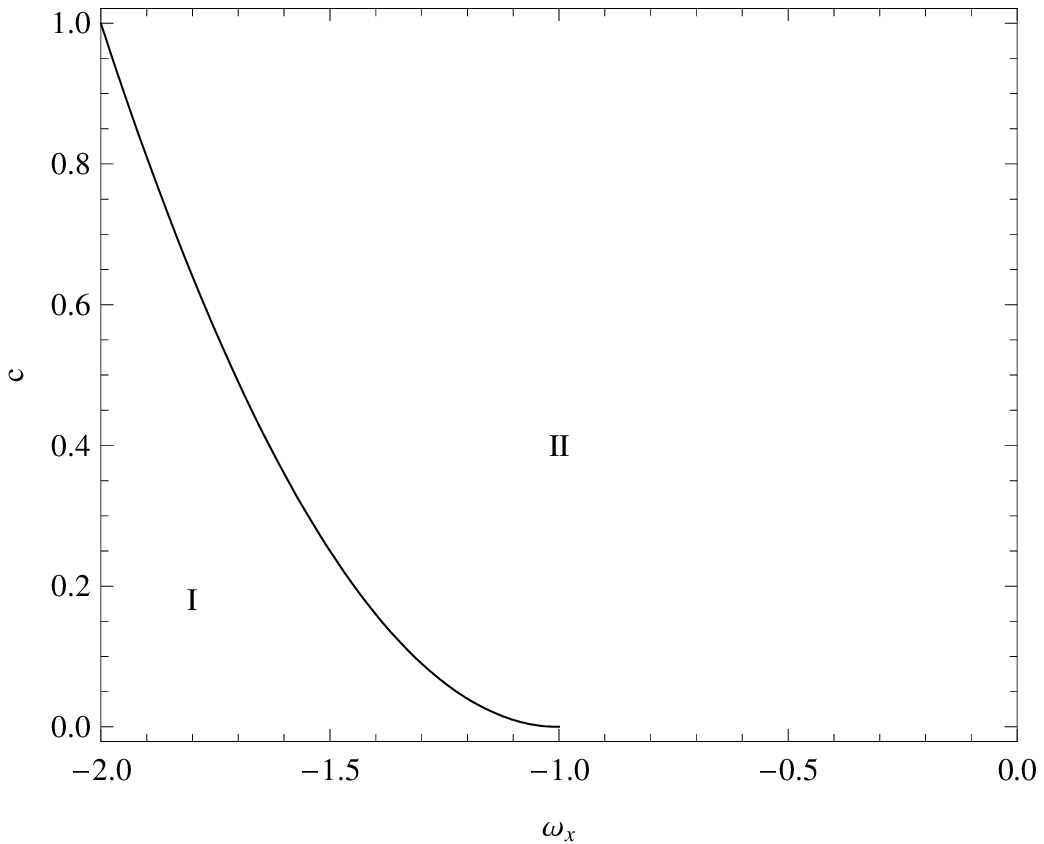}
\caption{The stable regions in the $(\omega_x, c)$ parameter space
for fixed $\alpha$ (the left for $\alpha=0.5$ and the right for
$\alpha=2$). In Einstein cosmology the critical points $A_2$
($\alpha=0.5$) and $B_4$ ($\alpha=2$) are late time attractor in the
region I+II. But in LQC, $A_2$ ($\alpha=0.5$) and $B_4$ are late
time attractor only in the region II. The region $III$ represents
the region of the solution without physical meaning. }
\end{center}
\end{figure}

\begin{figure}[ht]
\begin{center}
\includegraphics[width=6cm]{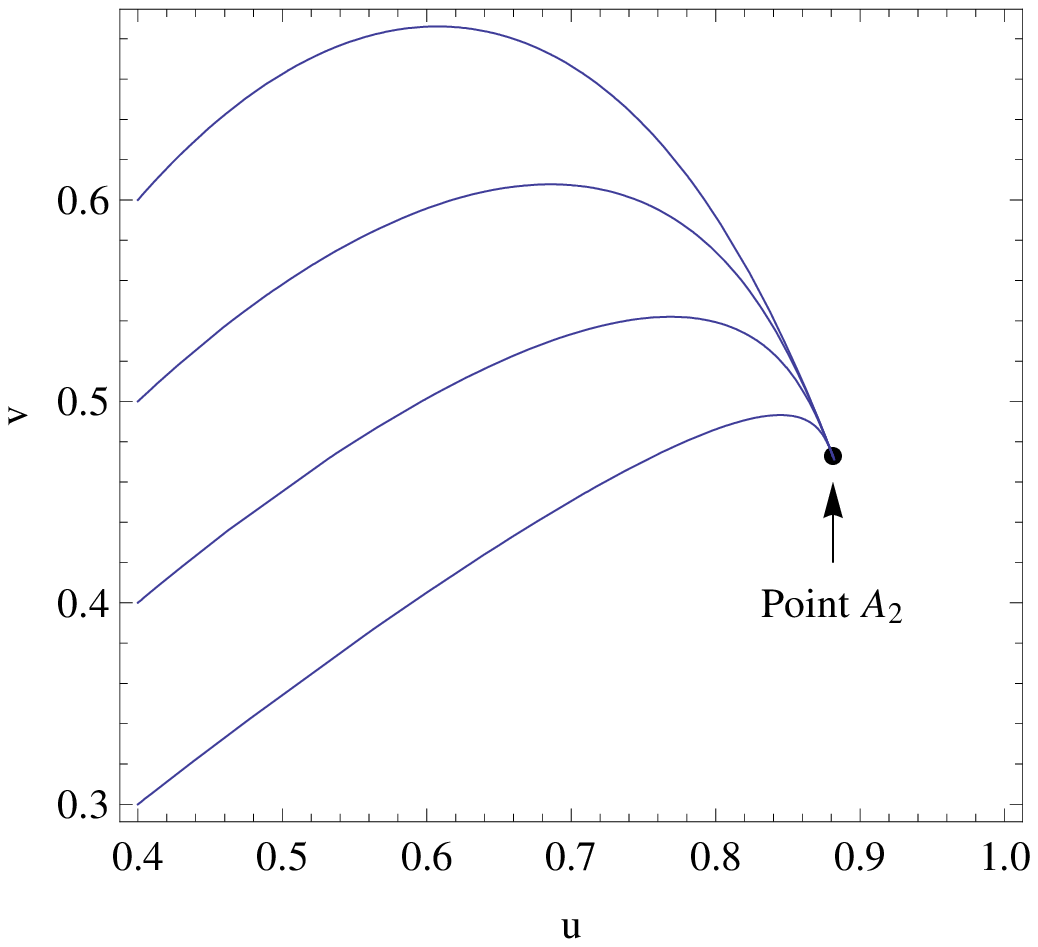}\;\;\;\;\;\includegraphics[width=6cm]{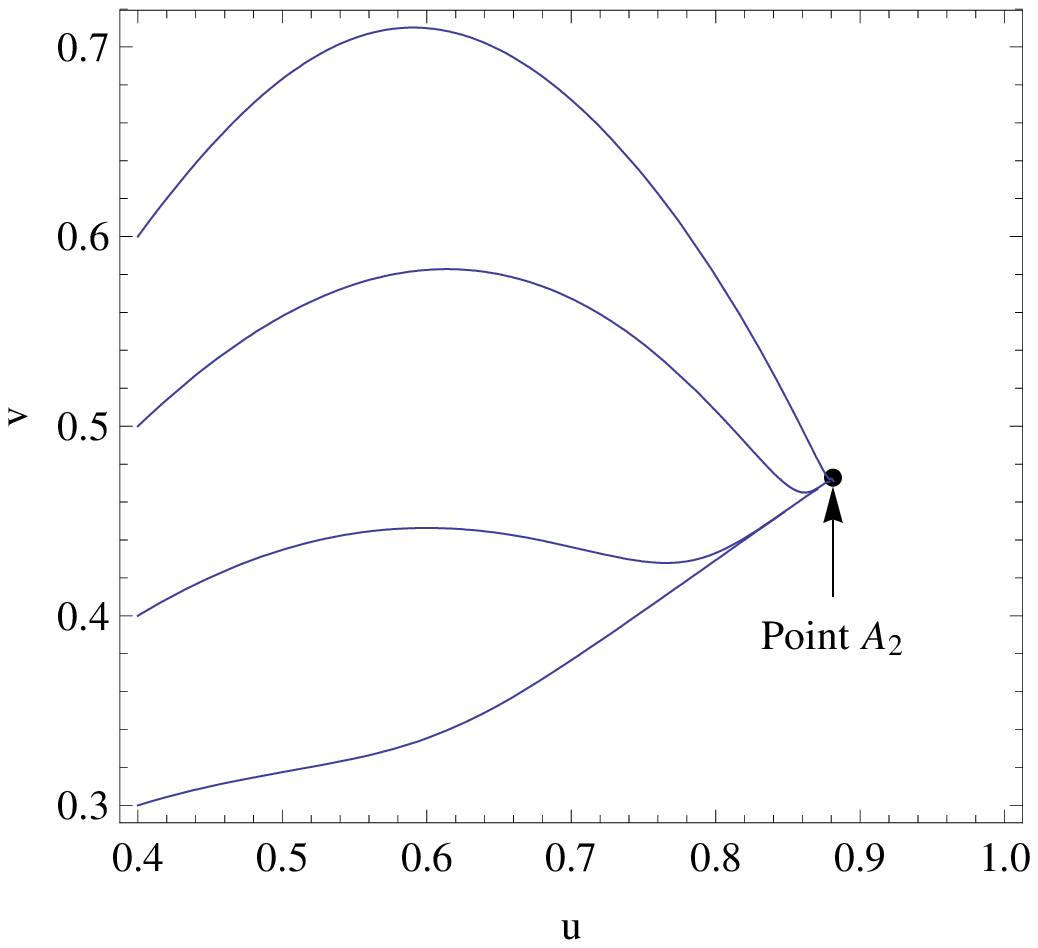}
\caption{The phase diagram of interacting dark energy (the left for
Einstein cosmology and the right for  LQC) with $\omega_x=-1.2$,
$c=0.5$ and $\alpha=0.5$. The point $A_2$ is the critical point.}
\end{center}
\end{figure}

\begin{figure}[ht]
\begin{center}
\includegraphics[width=6cm]{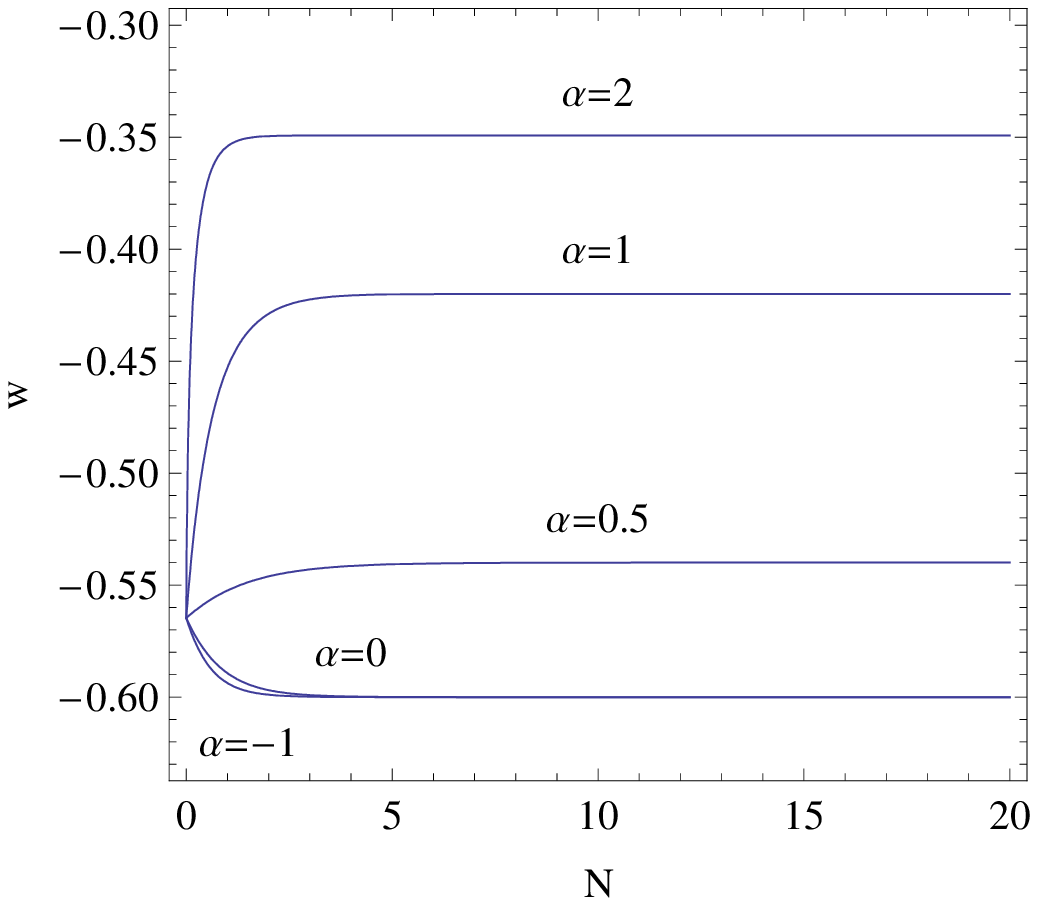}\;\;\;\;\;\includegraphics[width=6cm]{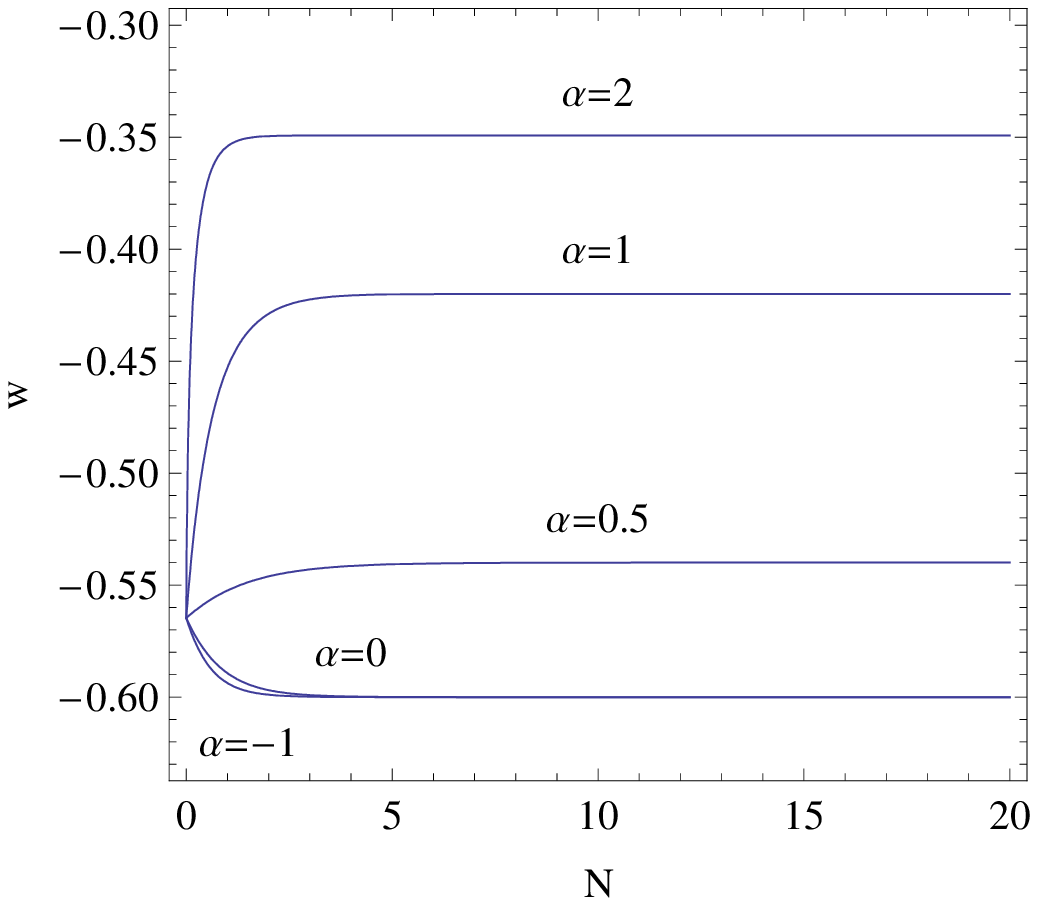}
\caption{The evolution of total cosmic energy $\omega$ for
$\omega_x=-0.6$, $c=0.18$ and fixed $\alpha$ (the left for Einstein
cosmology, and the right for LQC). }
\end{center}
\end{figure}

\begin{figure}[ht]
\begin{center}
\includegraphics[width=16cm]{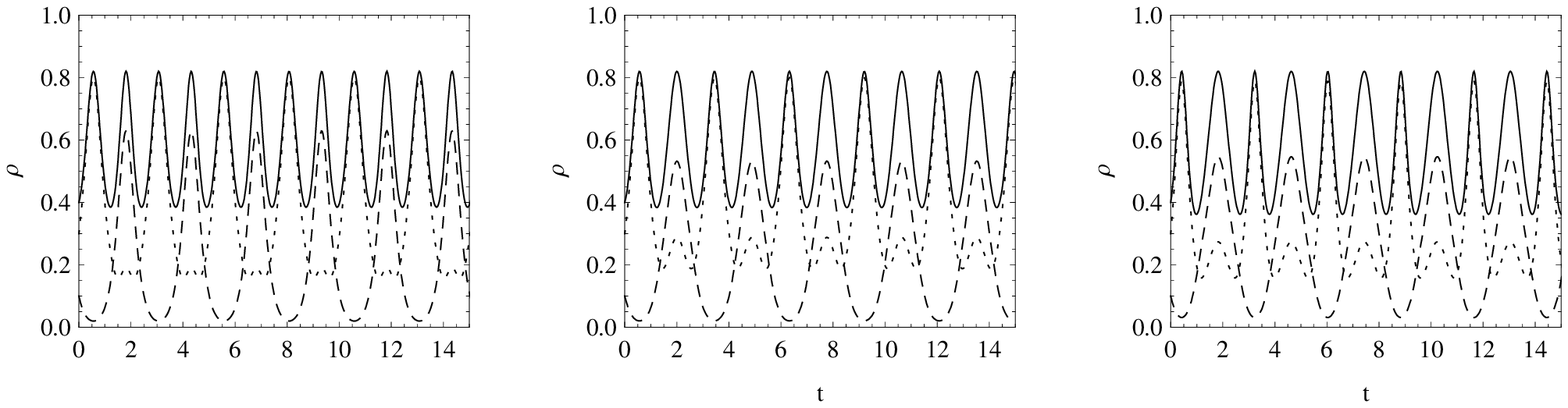}
\caption{The evolution of $\rho$ with $t$ in LQC for fixed
$\alpha=-1$ and $\rho_c=0.82$ (the left for $\omega_x=-1.6$ and
$c=0.1$, the middle for $\omega_x=-1.6$ and $c=0.3$, and the right
for $\omega_x=-1.8$ and $c=0.3$). The solid, dotted and dashed
curves correspond to $\rho_x+\rho_m$, $\rho_x$ and $\rho_m$,
respectively.}
\end{center}
\end{figure}


\vspace*{0.2cm}
\begin{thebibliography}{99}
\baselineskip=0.6 cm

\bibitem{A1}A. G. Riess et al, Astron. J. {\bf 116}, 1009 (1998); P. de Bernardis et. al.,
Nature {\bf 404}, 955 (2000);  A. G. Riess, et al.,
astro-ph/0611572.
\bibitem{A2} S. Perlmutter et al, Astrophys. J. {\bf 517}, 565 (1999); Astrophys. J. {\bf 598},
102 (2003).
\bibitem{A3} D. J. Eisenstein, et al., Astrophys. J. {\bf 633},
560 (2005).
\bibitem{A4} D. N. Spergel et al., Astrophys. J. Suppl.
{\bf 170}, 377 (2007).

\bibitem{T1} T. Padmanabhan, Phys. Rept. {\bf 380}, 235 (2003); P. J. E.Peebles, B.
Ratra, Rev. Mod. Phys. {\bf 75}, 559 (2003); V. Sahni, Lect Notes
Phys. {\bf 653}: 141 (2004) and references therein.


\bibitem{c1} W. Zimdahl, D. Pav¡äon, L.P. Chimento, Phys. Lett. B {\bf 521}, 133 (2001)
; L.P. Chimento, A. S. Jakubi, D. Pav¡äon, W. Zimdahl, Phys. Rev. D
{\bf 67}, 083513 (2003); S. del Campo, R. Herrera and D. Pav¡äon,
Phys. Rev. D {\bf 70},  043540 (2004); D. Pav¡äon and W. Zimdahl,
Phys. Lett. B {\bf 628}, 206 (2005).

\bibitem{c2}G. Olivares, F. Atrio-Barandela and D. Pav¡äon, Phys. Rev. D {\bf
71}, 063523 (2005); ibid. Phys. Rev. D {\bf 74}, 043521 (2006).

\bibitem{c3} B. Wang, Y. Gong and E. Abdalla, Phys. Lett. B {\bf 624}, 141 (2005);
B. Wang, Ch.-Y. Lin, Elcio Abdalla, Phys. Lett. B {\bf 637},
357(2006).

\bibitem{c4} B. Wang, J. Zang, Ch.-Y. Lin, E. Abdalla and S. Micheletti, Nucl.
Phys. B {\bf 778}, 69 (2007).

\bibitem{c5} S. Das, P.S. Corasaniti and J. Khoury, Phys. Rev. D {\bf 73},
083509 (2006).

\bibitem{c6} L. Amendola, Phys. Rev. D {\bf 62}, 042511 (2000); L. Amendola, D.
Tocchini-Valentini, Phys. Rev. D {\bf 64}, 043509 (2001); L.
Amendola, S. Tsujikawa and M. Sami, Phys. Lett. B {\bf 632}, 155
(2006); L. Amendola and C. Quercellini, Phys. Rev. D {\bf 68},
023514 (2003); G. W. Anderson and S. M. Carroll, astro-ph/9711288.

\bibitem{c7} W. Zimdahl, Int. J. Mod. Phys. D {\bf 14}, 2319 (2005).

\bibitem{c8} C. Feng, B. Wang, Y. Gong, R.-K. Su, JCAP {\bf 09}, 005 (2007).

\bibitem{O1} O. Bertolami, F. Gil Pedro and M. Le Delliou, Phys. Lett. B {\bf 654},
165 (2007). O. Bertolami, F. Gil Pedro and M. Le Delliou,
arXiv:0705.3118v1.

\bibitem{O2} E. Abdalla, L.Raul W. Abramo, L. Sodre Jr., B. Wang,
arXiv:0710.1198 [astro-ph].

\bibitem{pt} C. G. Bohmer, G. C. Cabral, R. Lazkoz, R. Maartens, Phys.Rev.{\bf D78}, 023505,(2008).

\bibitem{dM1} Z. K. Guo, N. Ohta and S. Tsujikawa, Phys. Rev. D {\bf 76}, 023508
(2007); L. Amendola, G. Campos, R. Rosenfeld, astro-ph/0610806.

\bibitem{dE1}D. Pavon, B. Wang, arXiv:0712.0565.

\bibitem{hb} J. He and B. Wang, JCAP 0806, 010, (2008); C. Feng, B. Wang, E. Abdalla, R.K.
Su, Phys.Lett.{\bf B 665} 111, (2008).

\bibitem{gf}G. Olivares, F. Atrio-Barandela and D. Pavon, Phys. Rev. D {\bf 77},
063513 (2008).

\bibitem{gf1} M. Manera and D. F. Mota,  Mon. Not. Roy. Astron. Soc. {\bf 371}, 1373
(2006),(astro-ph/0504519);  N. J. Nunes, and D. F. Mota, Mon. Not.
Roy. Astron. Soc. {\bf 368}, 751 (2006), (astro-ph/0409481).

\bibitem{lc1} M. Bojowald, Living Rev. Rel. {\bf 8}, 11 (2005); M.
Bojowald, gr-qc/0505057.

\bibitem{lc2} A. Ashtekar, M. Bojowald and J.
Lewandowski, Adv. Theor. Math. Phys. {\bf 7}, 233 (2003)
[gr-qc/0304074]; A. Ashtekar, gr-qc/0702030.

\bibitem{lc3} A. Ashtekar, AIP Conf. Proc. {\bf 861}, 3
(2006) [gr-qc/0605011].

\bibitem{lcg1} C. Rovelli, Living Rev. Rel. {\bf 1}, 1 (1998) [gr-qc/9710008]; T.
Thiemann, Lect. Notes Phys. {\bf 631}, 41 (2003) [gr-qc/0210094]; A.
Corichi, J. Phys. Conf. Ser. {\bf 24}, 1 (2005) [gr-qc/0507038]; A.
Perez, gr-qc/0409061.

\bibitem{lcg2} A. Ashtekar and J. Lewandowski, Class. Quant. Grav. {\bf 21}, R53
(2004) [gr-qc/0404018]; A. Ashtekar, arXiv:0705.2222 [gr-qc].

\bibitem{lcg3} C. Rovelli, Quantum Gravity, Cambridge University Press,
Cambridge (2004).  A. Ashtekar, New J. Phys. {\bf 7}, 198 (2005)
[gr-qc/0410054]; T. Thiemann, hep-th/0608210.

\bibitem{lcg31} A. Ashtekar, T. Pawlowski and P. Singh, Phys. Rev. Lett. {\bf 96},
141301 (2006) [gr-qc/0602086]; Phys. Rev. D {\bf 73}, 124038 (2006)
[gr-qc/0604013].

\bibitem{lcg32} P. Singh and A. Toporensky, Phys. Rev. D {\bf 69}, 104008 (2004) [gr-qc/0312110];
G. V. Vereshchagin, JCAP {\bf 0407}, 013 (2004) [gr-qc/0406108]; G.
Date and G. M. Hossain, Phys. Rev. Lett. {\bf 94}, 011302 (2005)
[gr-qc/0407074].

\bibitem{lcg33} M. Bojowald, Phys. Rev. Lett. {\bf 89}, 261301 (2002) [gr-qc/0206054];
M. Bojowald and K. Vandersloot, Phys. Rev. D {\bf 67}, 124023 (2003)
[gr-qc/0303072]; M. Bojowald, J. E. Lidsey, D. J. Mulryne, P. Singh
and R. Tavakol, Phys. Rev. D {\bf 70}, 043530 (2004)
[gr-qc/0403106].

\bibitem{lg4} A. Ashtekar, T. Pawlowski and P. Singh, Phys. Rev. D {\bf 74}, 084003
(2006) [gr-qc/0607039].

\bibitem{lg5} P. Singh, Phys. Rev. D {\bf 73}, 063508 (2006)
[gr-qc/0603043]; P. Singh, Class. Quant. Grav. {\bf 22}, 4203 (2005)
[gr-qc/0502086].

\bibitem{lg51} Y. Shtanov and V. Sahni, Phys. Lett. B {\bf 557}:1
(2003)[gr-qc/0208047].


\bibitem{lg6} M. sami, P. Singh and S. Tsujikawa, Phys. Rev. D {\bf 74},
043514 (2006); T. Naskar and J.Ward, arxiv: 0704.3606 [gr-qc].

\bibitem{lg7} D. Samart and B. Gumjupai, Phys. Rev. D {\bf 76}, 243514
(2007).

\bibitem{lg8} A. Ashtekar, A. Corichi and P. Singh, Phys. Rev. D {\bf 77}, 024046
(2008)[arxiv: 0710.3565]; A. Corichi and P. Singh,  Phys. Rev. D
{\bf 78}, 024034 (2008)[arxiv: 0805.0136].


\bibitem{phw1}P. X. Wu, S. N. Zhang, JCAP {\bf 06} (2008) 007.

\bibitem{phw2} X. Y Fu, H. W Yu and Puxun Wu,  Phys. Rev. D {\bf 78}, 063001 (2008).

\bibitem{phb1}B. Gumjudpai, arXiv:0706.3467

\bibitem{h1}H. Wei and S. N. Zhang,  Phys. Rev. D {\bf 76}, 063005 (2007).

\bibitem{pp} J.H. He, B. Wang, E. Abdalla, arXiv:0807.3471; J. Valiviita, E. Majerotto, R.
Maartens, arXiv:0804.0232; P. S. Corasaniti, arXiv:0808.1646.
\end{thebibliography}
\end{document}